\documentclass[
    prl,
    twocolumn,
    aps,
    superscriptaddress,
    nofootinbib,
    reprint,
]{revtex4-2}

\usepackage[T1]{fontenc}
\usepackage{amsmath,amssymb}
\usepackage{mathtools}

\usepackage[mathcal]{euscript}
\usepackage{microtype}
\usepackage{bm}

\usepackage{multirow}
\usepackage{graphicx}
\usepackage{subfig}
\graphicspath{{figures/}}
\usepackage{wrapfig}

\usepackage[svgnames,table]{xcolor}
\usepackage[
    colorlinks=true,
    citecolor=DimGray,
    linkcolor=SlateBlue,
    urlcolor=Navy,
    ]{hyperref}

\usepackage[capitalise]{cleveref}
\usepackage[
    range-phrase=\text{\textendash},
    range-units=single,
]{siunitx}
\DeclareSIUnit{\year}{yr}
\usepackage{dcolumn}

\usepackage{comment}
\usepackage{natbib}
\usepackage{orcidlink}

\newcommand{\rmd}{\mathrm{d}}
\newcommand{\rme}{\mathrm{e}}
\newcommand{\rmi}{\mathrm{i}}

\DeclareMathOperator{\sinc}{sinc}

\begin{document}

\title{First results from the Axion Dark-Matter Birefringent Cavity (ADBC) experiment}

\author{Swadha Pandey\,\orcidlink{0000-0002-2426-6781}}
\email{swadha@mit.edu}
\author{Evan D. Hall\,\orcidlink{0000-0001-9018-666X}}
\author{Matthew Evans\,\orcidlink{0000-0001-8459-4499}}
\affiliation{LIGO Laboratory, Department of Physics, Massachusetts Institute of Technology, Cambridge, MA 02139, USA}

\date{\today}

\begin{abstract}
Axions and axion-like particles are strongly motivated dark matter candidates that are the subject of many current ground based dark matter searches. We present first results from the Axion Dark-Matter Birefringent Cavity (ADBC) experiment, which is an optical bow-tie cavity probing the axion-induced birefringence of electromagnetic waves. Our experiment is the first optical axion detector that is tunable and quantum noise limited, making it sensitive to a wide range of axion masses.
We have iteratively probed the axion mass range $\qtyrange{40.9}{43.3}{\nano\eV}/c^2$, $\qtyrange{49.3}{50.6}{\nano\eV}/c^2$, and $\qtyrange{54.4}{56.7}{\nano\eV}/c^2$, and found no dark matter signal.
On average, we constrain the ALP--photon coupling at the level $g_{a\gamma\gamma} \leq \qty{1.9e-8}{\GeV^{-1}}$. We also present prospects for future axion dark matter detection experiments using optical cavities.
\end{abstract}

\maketitle

It is now more than forty years since the axion was proposed as a solution to the dark matter problem~\cite{Preskill:1982cy,Abbott:1982af,Sikivie:2020zpn}.
Since then, searches for ultralight dark matter have expanded from the canonical axion into a more general class of pseudoscalar axionlike particles (ALPs), whose mass $m_a$ could range from less than $\qty{e-20}{\eV}/c^2$ to $\qty{e-2}{\eV}/c^2$~\cite{Chadha-Day:2021szb}.

To arrive at an observable signature of the ALP, we need some additional hypotheses about the expected properties of dark matter.
Based on observations of Milky Way dynamics, the local density of dark matter is $\rho_\text{DM} = \qty{0.3}{\GeV/\cm^3}$;
additionally, dark matter is expected to be cold, with a typical velocity ${\sim}\num{e-3}c$~\cite{Read:2014qva}.
Together, these assumptions imply that for ALPs, the de Broglie wavelength is much larger than the typical interparticle spacing, and the dark matter is therefore well described as a classical field $a(t)$ oscillating near the ALP Compton frequency $\omega_a = m_a c^2 / \hbar$~\cite{Hui:2021tkt}.
Under the standard model of the Milky Way's dark matter halo, the occupation numbers follow a Maxwellian distribution in velocity~\cite{Freese:2012xd,Derevianko:2016vpm} (cf. Ref.~\cite{Evans:2018bqy}), meaning that the ALP field is Doppler broadened, with a fractional full-width half-maximum linewidth $\Delta\omega_a / \omega_a \approx \num{3e-6}$~\cite{Derevianko:2016vpm}.

The Lagrangian for interaction between an ALP field and an electromagnetic (EM) field with Faraday tensor $F_{\mu\nu}$ is given by
\begin{equation}
    \mathcal{L} \supset - \frac{1}{4} g_{a\gamma\gamma} a F_{\mu\nu}\Tilde{F}^{\mu\nu},
\end{equation}
where $g_{a\gamma\gamma}$ represents the strength of the ALP--photon coupling. This modifies Maxwell's equations, leading to electromagnetic signatures that are, in principle, observable~\cite{Sikivie:1983ip}.
One such signature is that photon-coupled ALPs induce circular birefringence between left-hand and right-hand polarized electromagnetic waves.
For such a wave wave at an angular frequency $\omega_0$ propagating through a classically oscillating ALP field $a(t)$ with Compton frequency $\omega_a \ll \omega_0$, the dispersion of the two circular polarization modes is~\cite{Liu:2018icu}
\begin{equation}
    c^2 k_\text{LCP,RCP}(t)^2 - \omega_0^2 = \pm\omega_0 \frac{g_{a\gamma\gamma}}{m_a} \sqrt{\frac{2\hbar^3 \rho_\text{DM}}{c}} \dot{a}(t),
    \label{eq:dispersion}
\end{equation}
with $k_{\text{LCP,RCP}}$ being the wavenumber of the left- or right-handed electromagnetic mode.
Searches for this signature have been proposed in the optical domain using resonant cavities~\cite{DeRocco:2018jwe,Obata:2018vvr,Liu:2018icu,Martynov:2019azm,Nagano:2019rbw,Nagano:2021kwx}, with initial experiments searching in the femto- to picoelectronvolt range~\cite{Oshima:2023csb} and around \qty{2}{\nano\eV}~\cite{Heinze:2023nfb}.
Searches for ALP-induced conversion between electromagnetic modes have also been proposed using superconducting rf cavities~\cite{Sikivie:2010fa,Berlin:2019ahk}.

\begin{figure}[t]
    \centering
    \includegraphics[width=\columnwidth]{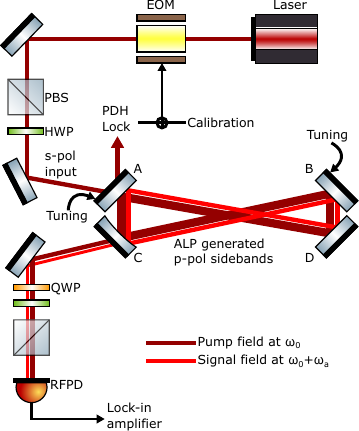}
    \caption{Experimental setup: bow-tie cavity with A, D, B, C mirrors. $\hat{s}$-polarized pump field from a 1064nm Nd:YAG laser enters the cavity at mirror A and is locked to the cavity using a PDH lock. ALP generated $\hat{p}$-polarized sidebands at the cavity splitting frequency $\omega_a=\omega_{\text{sp}}$ are resonant in the cavity. Heterodyne readout is performed using pump and signal field transmitted at mirror C.}
    \label{fig:setup}
\end{figure}
One of the experimental challenges of performing ALP searches with optical cavities is the ability to tune the cavity to search for ALPs of different masses.
In this work we have, for the first time, established a technique to demonstrate tunability in such a detector.
We thus present a search for ALPs near \qty{50}{\nano\eV} using a tunable and quantum-noise-limited birefringent optical cavity.

\emph{Experimental setup. --- }
Our experiment exploits the ALP-induced optical activity given by \cref{eq:dispersion}.
If the ALP field $a(t)$ is quasisinusoidal, with a central frequency at the Compton frequency $\omega_a$, then an $\hat{s}$-polarized electromagnetic wave with frequency $\omega_0$ propagating through the ALP field will develop $\hat{p}$-polarized phase sidebands at frequency $\omega_0 \pm \omega_a$\,---\,i.e., at a distinct frequency and orthogonal polarization to the pump field.
For propagation over a distance $\ell$, the amplitude of the $E^{(\omega_0\pm\omega_a,\hat{p})}$ sidebands relative to the $E^{(\omega_0,\hat{s})}$ pump amplitude is given by~\cite{Liu:2018icu}
\begin{equation}
    \beta_\ell \equiv \frac{E^{(\omega_0\pm\omega_a, \hat{p})}}{E^{(\omega_0, \hat{s})}} = \frac{g_{a\gamma\gamma}}{2\omega_a} \sqrt{c^3\hbar\rho_{\mathrm{DM}}} (e^{\pm\rmi\omega_a \ell/c} - 1).
    \label{eq:beta}
\end{equation}
This polarimetric rotation is of order \num{e-16} for $g_{a\gamma\gamma} = \qty{e-10}{\GeV^{-1}}$ and $\ell = \qty{1}{\meter}$.
To enhance this small signal, we constructed a birefringent bow-tie cavity as depicted in \cref{fig:setup}, using a \qty{1064}{\nm} Nd:YAG laser as a pump.
Our cavity is formed of four superpolished mirrors (labeled A, B, C, and D), with amplitude reflectivities $r$ satisfying $r_{D, B}^{\hat{p},\hat{s}} \gg r_{A, C}^{\hat{p},\hat{s}}$.
The mirror separations satisfy $L_{CA}, L_{DB} \ll L_{AD}, L_{BC}$, so that we take $L_{AD} \simeq L_{BC} \equiv L$.
Due to non-normal incidence, the mirrors cause a phase splitting between $\hat{s}$ and $\hat{p}$ polarized light upon reflection.
We have used mirrors such that this splitting is small for mirrors B and D, so the cumulative phase splitting $\Psi$ per cavity round-trip is dominated by mirrors A and C. 
When the carrier mode $(\omega_0,\hat{s})$ is resonant in the cavity, the cavity will also be resonant for the mode $(\omega_0 + \omega_{\text{sp}},\hat{p})$, where $\omega_{\text{sp}} = \Psi c/2L$ is the cavity frequency splitting between the $\hat{s}$ and $\hat{p}$ polarizations.
Thus for $\omega_{\text{sp}} = \omega_a$, the cavity is resonant for both the pump mode and for one of the two ALP-generated signal modes.

We lock our pump field to the cavity using a Pound–Drever–Hall lock~\cite{Drever:1983qsr}, with loop bandwidth of 80 kHz.
We characterize the cavity finesse in $\hat{s}$ by measuring the storage time~\cite{Isogai:2013wfa}, and in $\hat{p}$ by modulating the laser frequency to measure the cavity linewidth.
Since mirror transmission varies markedly with angle of incidence, transmission values were measured in situ for each cavity mirror in both polarizations.
The values for these parameters are given in \cref{tab:parameters} for our first dataset.
The cavity sits inside a steel enclosure on a floating table.

To be sensitive to a wide range of ALP frequencies, we must tune the cavity splitting.
We note that the reflection phase splitting at mirrors A and C is a sharp function of angle of incidence, and hence a small rotation of mirror B results in a significant shift in $\omega_{\text{sp}}$.
Mirror A is then rotated to close the cavity path again, and mirror C and the input optics are adjusted to maximize mode-matching and hence intra-cavity pump power for the new configuration.
This procedure is repeated each time to reach a new cavity splitting.

\begin{table}[t]
    \centering
    \caption{
        Experimental parameters: round-trip cavity length $2L$, finesse $\mathcal{F}$, input transmission $T_\text{A}$, output transmission $T_\text{C}$, input power $P_0$, laser wavelength $\lambda_0$, cavity splitting $\omega_{\mathrm{sp}}/2\pi$ for the first dataset. Where two values are given, the first refers to the $\hat{s}$-polarized pump mode, and the second refers to the $\hat{p}$-polarized signal mode.
        Values in parentheses denote uncertainties.
    \label{tab:parameters}}
    \begin{ruledtabular}
    \begin{tabular}{rSSl}
    {\textbf{Parameter}} & \multicolumn{2}{c}{\textbf{Value}} & {\textbf{Unit}} \\
    \hline
    $2L$ &
        \multicolumn{2}{S}{4.70(1)} & \unit{\meter} \\
    $\mathcal{F}$ &
        7260(70) & 212(1) & --- \\
    $T_\text{A}$ &
        5.3(1)e-4 & 0.0150(8) & --- \\
    $T_\text{C}$ &
        5.7(4)e-6 & 0.0130(3) & --- \\
    $P_0$ &
        \multicolumn{2}{S}{{0.8}} & \unit{\watt} \\
    $\lambda_0$ &
        \multicolumn{2}{S}{{1064}} & \unit{\nm} \\
    $\omega_{\mathrm{sp}}/2\pi$ &
        \multicolumn{2}{S}{{10.03}} & MHz \\
    \end{tabular}
    \end{ruledtabular}
\end{table}

At the transmission of mirror C, we perform a polarimetric heterodyne readout.
We use a quarter-wave plate to shift the signal modes from the phase quadrature to the amplitude quadrature relative to the pump field.
We then use a half-wave plate to project some of the transmitted pump field onto the same polarization state as the transmitted signal field.
Together, these two wave plates enable the production of an optical beat note at $\omega_\text{sp}$ if ALPs are present.
To sense this beat note, we use an rf photodiode with a bandwidth of 125 MHz.
We then send the ac output of the photodiode to a lock-in amplifier, where we demodulate the signal at the cavity splitting $\omega_{\textrm{sp}}$ over the signal mode cavity bandwidth, which is $\sim\qty{300}{\kHz}$.

For a photon shot noise limited measurement, we calculate the amplitude signal-to-noise ratio (see Supplementary Information):
\begin{multline}
\label{eq:gagg}
\text{SNR}
    = \frac{\epsilon g_{a\gamma\gamma} \sqrt{2P_{\text{cav}}^{\hat{s}}\rho_{\mathrm{DM}}\hbar c}t_C^{\hat{p}} \mathcal{F}_p L}{\sqrt{\hbar\omega_0}\pi} \\
    \times \left\lvert{\sinc{\left(\frac{\omega_a L}{c}\right)}}\right\rvert (\tau T)^{1/4},
\end{multline}
where $P_{\text{cav}}^{\hat{s}}$ is the intracavity pump power, $T$ is the integration time, $\tau = 1/\Delta\omega_a$ is the coherence time of the ALP field, $\mathcal{F}$ is the cavity finesse, $t_\text{C}^{\hat{p}}$ is the amplitude transmissivity of mirror C for the $\hat{p}$ polarization, and $\epsilon$ accounts for readout inefficiencies.

\emph{Calibration. --- }
To find the cavity splitting $\omega_{\text{sp}}$ for a given cavity configuration, we send linearly polarized light into the cavity at $\sim\qty{30}{\degree}$ relative to its eigenaxis and then lock the $(\omega_0,\hat{s})$ mode to the cavity.
We then drive an electro-optic modulator to generate $\hat{p}$ sidebands, which resonate when the sideband frequency matches $\omega_{\mathrm{sp}}$;
This appears as a beat note at the readout photodiode.

To calibrate the noise floor of our apparatus, we calculate the strength of signal produced by a phase fluctuation that mimics the axion background.
For a polarization rotation of $\beta_{2L}(\omega)$ over a single cavity round trip (\cref{eq:beta}), the demodulated ac power measured at the readout photodiode is
\begin{equation}
    P_{\mathrm{AC}}(\omega) = 2 \sqrt{P_{\mathrm{LO}} P_0} \mathcal{F}_s \mathcal{F}_p \beta_{2L}(\omega) t_C^{\hat{p}} t_A^{\hat{s}} \frac{\epsilon}{\pi^2} \left\lvert C^{\hat{p}}(\omega) \right\rvert,
    \label{eq:phase to power}
\end{equation}
where $C(\omega)$ is the normalized cavity amplitude transfer function.
We measure the amplitude spectral density (ASD) at the rf photodiode, referred to intracavity phase $\beta(\omega)$ using \cref{eq:phase to power} and our measured values of the finesses and transmissivities. (Details are provided in the Supplementary Information.)
We infer an intracavity phase sensitivity of $\qty{1e-12}{\radian/\sqrt{\Hz}}$, dominated by photon shot noise and with a \qty{-20}{\dB} contribution from electronics noise.

\begin{figure}[t]
    \centering
    \includegraphics[width=\columnwidth]{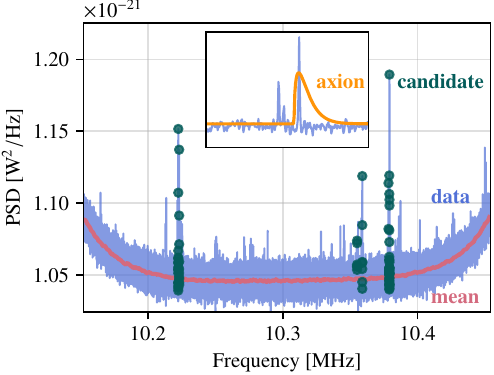}
    \caption{Mean-averaged PSD data (blue), neighboring-bin running median to estimate the mean (pink), and points that lie above the detection threshold (green) for the second dataset. We also show a portion of the data overlayed with the expected axion lineshape (orange), where it can be seen that the data peak is much narrower.}
    \label{fig:cand_search}
\end{figure}

\emph{Datataking and analysis. --- }
We took data in five discrete searches over the frequency range \qtyrange{9.88}{13.69}{\MHz}, with each dataset having a bandwidth of \qty{300}{\kHz}.
Each measurement was taken for 3 hours, with the data immediately demodulated at the cavity splitting frequency $\omega_\text{sp}$, Fourier transformed, and accumulated into a power spectral density (PSD) estimate, with $N \sim \num{47000}$ averages.
An example dataset is shown in \cref{fig:cand_search}.

For each dataset, we perform a search for an ALP signal in the PSD data.
A PSD estimate $S(\omega)$ resulting from $N$ mean-averaged periodograms converges to a Gaussian, with probability density
\begin{equation}
    P[S(\omega)] = \sqrt{\frac{N}{2\pi}}\frac{1}{\lambda(\omega)}
        \exp{\left\{-\frac{[S(\omega)-\lambda(\omega)]^2}{2\lambda(\omega)^2 / N}\right\}},
\end{equation}
with $\lambda(\omega)$ representing the underlying true value of the PSD.
For the case of no ALP signal\,---\,i.e., the null hypothesis $\mathcal{H}_0$\,---\,the value of $\lambda(\omega)$ is just given by the detector noise PSD, which we call $\lambda_0(\omega)$.
We infer $\lambda_0(\omega)$ using a running median of the neighboring 500 bins in the data.
Since an axion signal is expected to be ${\sim}10$ bins wide, this gives us a background-only estimate.

In the presence of an ALP at $\omega_a$, the ALP signal would have a power spectral density $\Gamma_{\omega_a}(\omega) F_{\omega_a}(\omega)$, where $F_{\omega_a}(\omega)$ is the ALP line-shape~\cite{Derevianko:2016vpm}, and $\Gamma_{\omega_a}(\omega) \propto g_{a\gamma\gamma}^2$.
Thus the total fluctuation in the detector has PSD
\begin{equation}
    \lambda(\omega) = \lambda_0(\omega) + \Gamma_{\omega_a}(\omega) F_{\omega_a}(\omega).
\end{equation}
For each potential ALP frequency, we construct an optimal test statistic
\begin{equation}
    Y(\omega) = \left(\sum_{\omega^\prime}\frac{F_{\omega}(\omega^\prime)^2}{\lambda_0(\omega^\prime)^2}\right)^{-1} \sum_{\omega^\prime}\frac{F_{\omega}(\omega^\prime)}{\lambda_0(\omega^\prime)^2} [S(\omega^\prime)-\lambda_0(\omega^\prime)],
\end{equation}
with uncertainty $\sigma_Y(\omega) = \left[N\sum_{\omega'} F_\omega(\omega')^2 / \lambda_0(\omega^\prime)^2 \right]^{-1/2}$.
Under the null hypothesis $\mathcal{H}_0$, $Y(\omega)/\sigma_Y(\omega)$ follows the standard normal distribution; otherwise, if an ALP is present with mass $m_a$ and coupling $g_{a\gamma\gamma}$, then $Y(\omega)$ converges to $\Gamma_{\omega_a}(\omega)$.
To reject $\mathcal{H}_0$ at the $5\sigma$ level, we would need a nonzero value of $Y$ with an overall significance $\alpha = \num{2.9e-7}$.
We therefore seek a threshold value $Y_*^{(n)}(\omega)$ satisfying
\begin{equation}
    P{\left(Y(\omega) > Y_\ast^{(n)}(\omega) \,\middle\vert\, \mathcal{H}_0\right)} = \alpha/n,
\end{equation}
with $n$ being the number of independent tests on the data~\cite[\S10.7]{Wasserman}; i.e., the total bandwidth of our data divided by the ALP linewidth.
For our data, $n = \num{8287}$ and hence $Y_\ast^{(n)} (\omega) = 6.5\sigma_{Y}(\omega)$.
We search for any data with $Y(\omega) > Y_*^{(n)}(\omega)$, and find tens to hundreds of such candidate points per dataset. 
This is shown in \cref{fig:cand_search} for dataset 2.
We now use our knowledge of the fractional full-width half-maximum linewidth of the ALP signal, $\Delta\omega_a / \omega_a \approx \num{3e-6}$~\cite{Derevianko:2016vpm}. To each candidate we fit a Lorentzian lineshape via least squares regression and find that no candidate has a linewidth within a factor of two of the expected ALP linewidth (most of the lines are found to be too narrow).
We therefore reject all candidates and conclude that no ALP signal is present in the data.

\begin{figure}[t]
    \centering
    \includegraphics[width=\columnwidth]{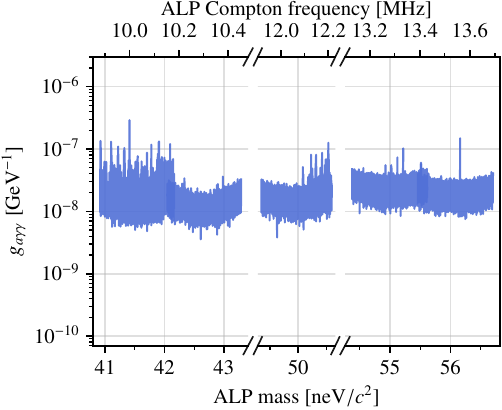}
    \caption{95\% upper limit on $g_{a\gamma\gamma}$ placed by the first run of the ADBC experiment. We have bounds from five datasets over axion frequency ranges \qtyrange{9.88}{10.45}{\MHz}, \qtyrange{11.92}{12.22}{\MHz}, and \qtyrange{13.12}{13.69}{\MHz} with an average sensitivity of \qty{1.9e-8}{\giga\eV^{-1}}.}
    \label{fig:bounds}
\end{figure}

We proceed to set upper bounds on $g_{a\gamma\gamma}$ at $\omega$ at the \qty{95}{\%} upper limit by using the distribution of $Y(\omega)$ under $\mathcal{H}_1(g_{a\gamma\gamma}^{95\%}, \omega_a = \omega)$, the hypothesis for the existence of an ALP field with rest-frame Compton frequency $\omega_a = \omega$ and coupling strength $g_{a\gamma\gamma}^{95\%}$.
This distribution is normal with
\begin{equation}
\label{eq:frequentistcl}
    P{\left(Y(\omega)  \,\middle\vert\, \mathcal{H}_1(g_{a\gamma\gamma}^{95\%}, \omega_a = \omega)\right)} = \mathcal{N}{\left(\Gamma_{\omega_a=\omega}(\omega), \sigma_{Y}(\omega)^2\right)}
\end{equation}
Since $Y(\omega)$ can take negative values, we use the Feldman--Cousins approach~\cite{Feldman:1997qc} to ensure non-negative confidence intervals. 
The upper limits thus obtained from our data are shown in \cref{fig:bounds} for the 5 datasets.
The average sensitivity we have achieved is $g_{a\gamma\gamma}^{95\%} \leq \qty{1.9e-8}{\giga\eV^{-1}}$ over the probed frequency range.

\emph{Appraisal and future upgrades. --- }
We have performed an ALP dark matter search using a \qty{5}{\meter} optical bow-tie cavity over five different ALP masses in the range $\qtyrange{40.9}{56.7}{\nano\eV}/c^2$, corresponding to an ALP Compton frequency $\qtyrange{9.88}{13.69}{\MHz}$. 
Each search had a sensitivity band of \qty{300}{\kHz} and we have probed the ALP-photon coupling at an average sensitivity of $g_{a\gamma\gamma}^{95\%} \leq \qty{1.9e-8}{\giga\eV^{-1}}$ over all datasets. 
In this process, we have demonstrated for the first time an optical polarimetry based ALP detector whose search range has been enhanced by frequency tunability.

\begin{figure*}[t]
    \centering
    \includegraphics[width=0.9\textwidth]{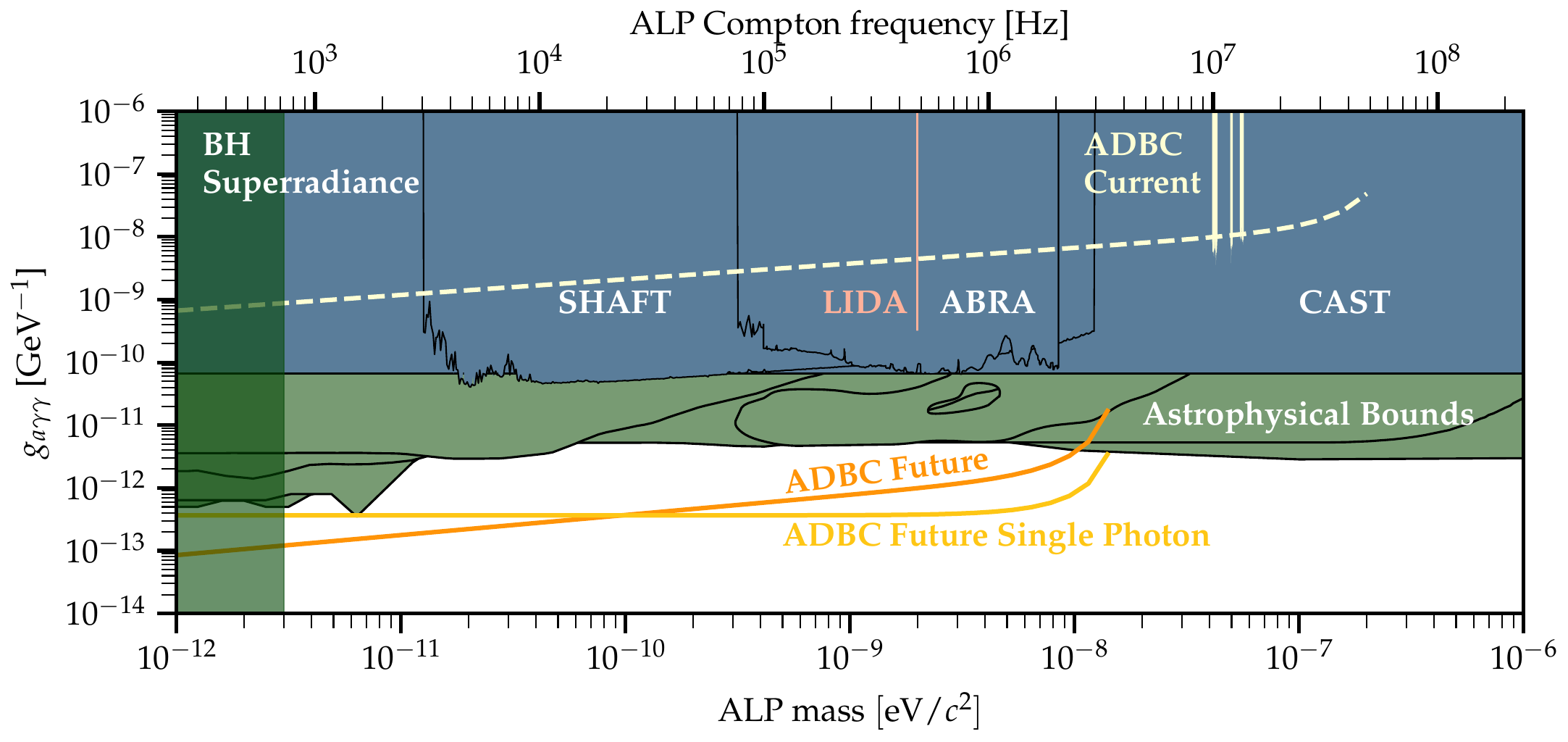}
    \caption{
    Current bounds and future projections from implementations of optical ALP polarimetry.
    We show ADBC's current data run, along with a dashed line indicating the apparatus's current sensitivity if we performed a search over the full mass range.
    Sensitivities are also shown for a future ADBC upgrade still using heterodyne readout, as well as the same apparatus operated with single photon readout. Current bounds from other ALP polarimetry experiments include LIDA (shown in plot) and DANCE ($g_{a\gamma\gamma} \leq \qty{8e-4}{\giga\eV^{-1}}$ for $\qty{e-14}{\eV} < m_a c^2 < \qty{e-13}{\eV}$).
    The blue regions show bounds from terrestrial ALP searches, in which we highlight the bounds from the solar axion search CAST~\cite{CAST:2007jps}, and the toroidal magnet searches ABRACADABRA~\cite{Salemi:2021gck} and SHAFT~\cite{Gramolin:2020ict}.
    The green regions show various astrophysical constraints, particularly constraints from black hole superradiance~\cite{AxionLimits}. 
    \label{fig:future bounds}
    }
\end{figure*}

A direct upgrade to this experiment would involve higher intracavity power, lower mirror transmission, and building a longer cavity.
Automation of the tuning process would also be required in order to scan the entire free spectral range of the cavity.
To reach shot-noise-limited sensitivity in the kilohertz band and below, more aggressive vibrational isolation may be required.
Projections for an upgrade with $P_{\mathrm{cav}} = \qty{1}{\mega\watt}$, $\mathcal{F}_p = 10^5$, and $L = \qty{40}{\meter}$ are shown in \cref{fig:future bounds}.

Photon counting for ALP detection has been proposed as an alternative~\cite{Lamoreaux:2013koa,Yu:2023wsn} to heterodyne readout, where phase information is sacrificed for enhanced sensitivity.
In the optical domain, this involves filtering out the pump photons at the readout port and using a single photon detector to measure the presence of any signal photons exiting the cavity.
Assuming Poissonian statistics for the photon fields, the SNR for such a measurement scheme is
\begin{equation}
    \text{SNR} = \frac{\Dot{N}_p T}{\sqrt{(2(\Dot{N}_s + \Dot{N}_\text{dark}) + \Dot{N}_p) T}},
\end{equation}
where $\Dot{N}_p$ and $\Dot{N}_s$ are the rates of signal ($\hat{p}$) and pump ($\hat{s}$) photons reaching the single photon detector, and $\Dot{N}_\text{dark}$ is the dark count rate.
To avoid being dominated by pump photon or dark noise at the detector, one has to make an optimistic projection of a dark count rate of 1 per hour and an extinction ratio of $\sim10^{23}$ of pump to signal photons.
One possible avenue includes a series of frequency-selective optical cavities located at the main cavity readout port, tuned to pass photons at the signal frequency and reject photons at the pump frequency~\cite{McCuller:2022hum}.
The fact that the pump and signal modes are generated in orthogonal polarizations also enables some pump filtering via polarization-selective optics. 
Projections for the ADBC upgrade using this photon counting scheme are shown in \cref{fig:future bounds}.
This assumes a total integration time $T_\text{total} = \qty{1}{\year}$, allocated equally among all the dwell frequencies.

\cref{fig:future bounds} also shows expected constraints from gravitational-wave searches for superradiant generation of boson fields around stellar-mass black holes~\cite{Brito:2017zvb}.
The stringency of these searches depends on the abundance of highly spinning black holes, which are susceptible to superradiant instability.
These searches may directly reveal gravitational-wave emission from superradiant clouds~\cite{Brito:2017wnc}, or the presence of such clouds may be inferred from the nonobservation of highly spinning black hole systems at a particular mass~\cite{Ng:2019jsx}.
Under the conservative assumption that stellar-mass black holes are born with low spin, the current generation of gravitational-wave detectors (LIGO--Virgo--Kagra) could infer the existence of ALPs with mass \qtyrange{1e-13}{3e-12}{\eV} after accumulating several hundred binary black hole merger observations~\cite{Ng:2019jsx}.
A similar constraint under conservative low-spin assumptions could be attained by searches for a stochastic background of gravitational waves with the next generation of gravitational-wave detectors, scheduled to come online in the 2030s~\cite{Yuan:2021ebu}.

\emph{Acknowledgments. --- }
SP was supported by the Bruno Rossi Graduate Fellowship in Astrophysics and the Charles E. Ross Fund.
EDH was supported by the MathWorks, Inc.
The authors thank Myron MacInnis for technical support, and Nancy Aggarwal, Rainer Weiss, and Lisa Barsotti for comments on the manuscript.

\appendix
\section{Supplementary Information for ``First results from the Axion Dark-Matter Birefringent Cavity (ADBC) experiment''}

\section{Sensitivity calculation}
\label{sec:sensitivity_calculation}

\begin{figure}[t]
    \centering
    \includegraphics[width=\columnwidth]{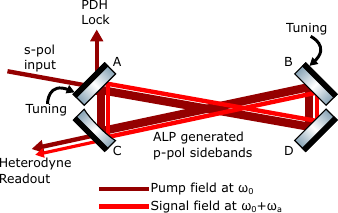}
    \caption{Experimental setup: bow-tie cavity with A, D, B, C mirrors. $\hat{s}$-polarized pump field enters the cavity at mirror A and is locked to the cavity using a PDH lock. ALP generated $\hat{p}$-polarized sidebands at the caity splitting frequency $\omega_a=\omega_{\text{sp}}$ are resonant in the cavity. Heterodyne readout is performed using pump and signal field transmitted at mirror C.}
    \label{fig:setup_supp}
\end{figure}

\subsection{Generation of polarization rotation by ALPs}

The Lagrangian for interaction between a pseudoscalar ALP field $a(t) = a_0 \cos{\omega_a}t$ at the axion frequency $\omega_a$ and an electromagnetic (EM) field with Faraday tensor $F_{\mu\nu}$ is given by
\begin{equation}
    \mathcal{L} \supset - \frac{1}{4} g_{a\gamma\gamma} a F_{\mu\nu}\Tilde{F}^{\mu\nu}.
\end{equation}
This modifies Maxwell's equations and results in the axion induced dispersion relation between an EM field of left circular polarization (LCP) and right circular polarization (RCP):
\begin{equation}
    k_{\mathrm{LCP, RCP}} (t) 
    = \frac{\omega_0}{c} \pm \frac{g_{a\gamma\gamma}}{2m_a} \sqrt{\frac{\hbar^3 \rho_{\mathrm{DM}}}{c^3}} \Dot{a}(t),
\end{equation}
where $m_a$ is the axion mass, $\rho_{\mathrm{DM}}$ is the local dark matter density, $\omega_0$ is the EM field angular frequency. 
Then the phase difference between the two polarizations upon propagation for a distance $\ell$ is
\begin{equation}
    \Delta \phi (t) = \phi_{\mathrm{LCP}} (t) - \phi_{\mathrm{RCP}} (t)
    = \int\limits_0^{L} \rmd{}x\, \Delta k{\left(t - \frac{x}{c}\right)}.
\end{equation}
Taking a Laplace transform and substituting $s = i\Omega$ yields
\begin{equation}
    \Delta \phi (\Omega) = \frac{g_{a\gamma\gamma}}{m_a} \sqrt{\frac{\hbar^3 \rho_{\mathrm{DM}}}{c}} a(\Omega) (\rme^{\rmi\Omega{L}/c} - 1).
\end{equation}
We can switch back to the time domain by assuming, for the moment, that $a$ is nearly monochromatic, so that $a(\Omega) \simeq \tfrac{1}{2}[\delta(\Omega-\omega_a) + \delta(\Omega+\omega_a)]$ (i.e., $a(t) = a_0 \cos{\omega_a t}$).
Then
\begin{equation}
  \Delta \phi (t) = \frac{g_{a\gamma\gamma}}{\hbar\omega_a} \sqrt{c^3\hbar^3\rho_{\mathrm{DM}}} \left(\rme^{\rmi\omega_a{L}/c} - 1\right)\frac{\rme^{\rmi\omega_a t}}{2} + \text{cc}.
\end{equation}
Thus with the complex electric field amplitude $\mathbf{E}$ of the EM wave expressed in the circular polarization basis\,---\,i.e., by the two-element vector $\left(\begin{smallmatrix} \mathbf{E} \cdot \hat{\mathbf{e}}_\text{LCP}\\ \mathbf{E} \cdot \hat{\mathbf{e}}_\text{RCP} \end{smallmatrix}\right)$\,---\,the propagation matrix for the amplitude over a length ${L}$ in the presence of an ALP field is
\begin{equation}
    P_{L} = 
    \begin{pmatrix}
        \rme^{-i\Delta\phi(t)/2} & 0\\
        0 & \rme^{i\Delta\phi(t)/2}
    \end{pmatrix}.
\end{equation}
In the linearly polarized basis for the electric field amplitude, with elements $\hat{\mathbf{e}}_s$ and $\hat{\mathbf{e}}_p$, the propagation matrix is instead
\begin{equation}
    P_{L} = 
    \begin{pmatrix}
        \cos{[\Delta\phi(t)/2]} & -\sin{[\Delta\phi(t)/2]}\\
        \sin{[\Delta\phi(t)/2]} & \cos{[\Delta\phi(t)/2]}
    \end{pmatrix}.
\end{equation}

\subsection{Cavity enhancement of the ALP-generated polarization signal}

For this experiment, we assume that the pump mode is $(\omega_0,\hat{s})$, and the resonant signal mode is $(\omega_0 + \omega_a,\hat{p})$; i.e., these are the resonant frequencies of the bow-tie cavity as shown in \cref{fig:setup_supp} and so we only need to consider these two modes for our analysis.
(The same analysis applies if $(\omega_0 - \omega_a, \hat{p})$ were resonant.)
Thus rewriting the propagation matrix with the two basis elements $\hat{\mathbf{e}}_{(\omega_0,s)}$ and $\hat{\mathbf{e}}_{(\omega_0 + \omega_a,p)}$ allows us to absorb the time-dependent factor into the basis itself:
\begin{equation}
    P_{L} \approx 
    \begin{pmatrix}
        \rme^{\rmi\omega_0 {L}/c} & 0\\
        \beta_L & \rme^{\rmi(\omega_0+\omega_a){L}/c}
    \end{pmatrix},
\end{equation}
where
\begin{equation}
\label{eq:beta_app}
    \beta_L = \frac{g_{a\gamma\gamma}}{2\hbar\omega_a} \sqrt{c^3\hbar^3\rho_{\mathrm{DM}}} (e^{\rmi\omega_a  L/c} - 1),
\end{equation}
and we have assumed $\beta_L \ll 1$. The input field is given by
\begin{equation}
    E_{\mathrm{in}} = E_0
    \begin{pmatrix}
        1\\
        0
    \end{pmatrix}
\end{equation}
and for the setup given in \cref{fig:setup}, we have reflectivities of mirrors $r_{D, B}^{\hat{p},\hat{s}} \gg r_{A, C}^{\hat{p},\hat{s}}$. 
We also have $L_{CA}, L_{DB} \ll L_{AD}, L_{BC} = L$ and that the phase splitting per cavity reflection $\Psi$ between $\hat{p}$ and $\hat{s}$ polarized light comes solely from mirrors A and C. 
Then the cavity round-trip reflection matrix is given by
\begin{equation}
    r = 
    \begin{pmatrix}
        r_A^{\hat{s}}r_C^{\hat{s}} && 0\\
        0 && r_A^{\hat{p}}r_C^{\hat{p}}e^{i\Psi}
    \end{pmatrix}
\end{equation}
and the sequence of bounces implies a round-trip propagation matrix $r P_L^2$.
We can therefore solve for the cavity field using
\begin{equation}
    E_{\mathrm{cav}} = t_A^{\hat{s}} E_{\mathrm{in}} + r P_L^2 E_{\mathrm{cav}}.
\end{equation}
Solving the above equation with resonance condition for the cavity $\hat{s}$-$\hat{p}$ phase splitting $\Psi = -2L\omega_a/c$ gives
\begin{equation}
    E_{\mathrm{cav}}^{(\omega_0 + \omega_a, \hat{p})} = 
    \frac{-G^\prime e^{i\Psi} t_A^{\hat{s}} E_0 (e^{2i\omega_a L/c}-1)}{(1-r_A^{\hat{p}}r_C^{\hat{p}})(1-r_A^{\hat{s}}r_C^{\hat{s}})},
\end{equation}
where $G^\prime = (g_{a\gamma\gamma}/2\hbar\omega_a) \sqrt{c^3\hbar^3 \rho_{\mathrm{DM}}}$.
We can define $\mathcal{F}^{\hat{p}} = \pi / (1 - r_A^{\hat{p}} r_C^{\hat{p}})$, and similarly for $\mathcal{F}^{\hat{s}}$, and we can define $\beta_{2L}(\omega) = G' (\rme^{2\rmi\omega_a L/c} - 1)$.
The output field from mirror C is then given by
\begin{align}
  E_{\mathrm{out}}^{(\omega_0 + \omega_a,\hat{p})} &= \rme^{2\rmi(\omega_0+\omega_a)L/c}t_C^{\hat{p}} E_{\mathrm{cav}}^{(\omega_0 + \omega_a,\hat{p})} \\
  &= -\rme^{2\rmi(\omega_0+\omega_a)L/c} t_C^{\hat{p}} t_A^{\hat{s}} \rme^{\rmi\Psi} \beta_{2L}(\omega) \frac{\mathcal{F}^{\hat{p}} \mathcal{F}^{\hat{s}}}{\pi^2} E_0. \label{eq:Eout_p}
\end{align}
\cref{eq:Eout_p} assumes that $\omega_a$ is precisely coincident with the signal mode ($\hat{p}$) resonance frequency $\omega_{\text{res}}^{\hat{p}}$; otherwise, the expression must be multiplied by
\begin{equation}
  C^{\hat{p}}(\omega_a) = \frac{1}{1 + \rmi(\omega_a - \omega_{\text{res}}^{\hat{p}}) / \omega_{\text{pole}}^{\hat{p}}},
  \label{eq:Chat_p}
\end{equation}
which is the normalized cavity frequency response for the signal mode, with $\omega_{\text{pole}}^{\hat{p}}$ being the cavity pole.

\subsection{Polarimetric readout of the cavity}

The heterodyne readout scheme is shown in \cref{fig:calibration}. The quarter-wave plate is used to bring the transmitted pump and signal fields into the same quadrature. 
We then use the half-wave plate to convert some of the transmitted pump field into a $\hat{p}$-polaried local oscillator field $E_{\text{LO}}^{(\omega_0, \hat{p})}$ at frequency $\omega_0$.
Since the signal field \cref{eq:Eout_p} oscillates at $\omega_a$ relative to the local oscillator, an rf beat note developes, with instantaneous power
\begin{equation}
  P_{\text{ac}}(t) = \tfrac{1}{2} c \varepsilon_0 \left[{E_{\text{LO}}^{(\omega_0,\hat{p})}}^* E_{\text{out}}^{(\omega_0+\omega_a,\hat{p})} \rme^{\rmi\omega_a t} + \text{cc} \right].
  \label{eq:Pac}
\end{equation}
The time-averaged power in the beat note is $2\sqrt{P_{\text{LO}}^{(\omega_0,\hat{p})} P_{\text{out}}^{(\omega_0+\omega_a,\hat{p})}}$.

For shot noise limited measurement, amplitude spectral density (ASD) of the noise is given by
\begin{equation}
    N = \sqrt{2\hbar\omega_0 P_{\text{LO}}^{(\omega_0, \hat{p})}}.
\end{equation}
giving a signal-to-noise ratio (SNR)
\begin{equation}
    \mathrm{SNR} = \frac{2\sqrt{P_{\text{LO}}^{(\omega_0, \hat{p})}P_{\mathrm{out}}^{(\omega_0+\omega_a, \hat{p})}}(\tau T)^{1/4}}{\sqrt{2\hbar\omega_0 P_{\text{LO}}^{(\omega_0, \hat{p})}}}
\end{equation}
where $T$ is the integration time and $\tau$ is the coherence time of the ALP field. 
Ref.~\cite{Derevianko:2016vpm} calculates the expected ALP lineshape, which is
\begin{equation}
\label{eq:F}
    F_{\omega_a}(\omega) \propto e^{-(\omega-\omega_a)\tau} \sinh{\left[\eta\sqrt{\eta^2 + 2(\omega-\omega_a)\tau}\right]}.
\end{equation}
Here $\tau \equiv c^2/(v^2 \omega_a)$, with $v$ being related to the most probable speed $v_\text{most}$ in the standard halo model by $v_\text{most} = \sqrt{2} v$, and $\eta = v_\text{gal} / v$~\cite{Freese:2012xd,Derevianko:2016vpm}.
Then for mode-matching and readout efficiency factor $\epsilon$, cavity finesse $\mathcal{F}$, and normalized cavity amplitude transfer function $C(\omega)$, the expected PSD to be measured at the readout is given by
\begin{multline}
\label{eq:s_sig}
    \langle S_\text{sig} (\omega,\omega_a,g_{a\gamma\gamma})\rangle 
    = F_{\omega_a}(\omega) \times \left[2 g_{a\gamma\gamma}\epsilon \left\lvert C^{\hat{p}}(\omega) \right\rvert \right]^2\\
    \times \left[\sqrt{2P_0 P_{\text{LO}}\rho_{\mathrm{DM}}\hbar c}L t_C^{\hat{p}} t_A^{\hat{s}} \frac{\mathcal{F}_s \mathcal{F}_p}{\pi^2} \right. \\
    \left.\times\left|\sinc{\left(\omega_a L/c\right)}\right|(\tau T)^{1/4}\right]^2
\end{multline}
with $C^{\hat{p}}$ defined as in \cref{eq:Chat_p}.

\begin{figure}[t]
    \centering
    \includegraphics[width=\columnwidth]{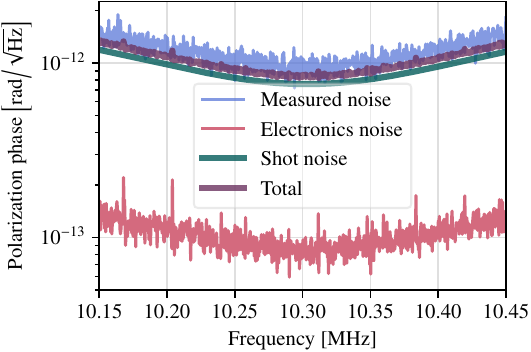}
    \caption{Polarimetric phase sensitivity of the apparatus for the second dataset, with contributions from shot noise calculated from measured power (green), and measured photodiode dark noise (pink). Total measured noise (blue) is $13\%$ above the total modeled noise (purple). The error bars on the total modeled noise are calculated to be $6.5\%$.}
    \label{fig:noise budget}
\end{figure}

\section{Noise calibration}
\label{sec:noise}

To calibrate the noise floor of our apparatus, we measure the ASD at the demodulated ac output of the rf photodiode.
This can be converted into an equivalent axion-induced intracavity polarization phase rotation $\beta_{2L}(\omega)$ over a single cavity round trip by computing the absolute value of the transfer function from $\beta$, given by \cref{eq:beta_app}, into $P$, given by \cref{eq:Pac}:
\begin{multline}
    \bigl\lvert P_{\text{AC}}(\omega) \bigr\rvert
    = 2 \sqrt{P_{\mathrm{LO}}^{(\omega_0, \hat{p})} P_0} \mathcal{F}_s \mathcal{F}_p t_C^{\hat{p}} t_A^{\hat{s}} \frac{\epsilon}{\pi^2} \left\lvert \beta_{2L}(\omega) C^{\hat{p}}(\omega)\right\rvert.
      \label{eq:beta_calib}
\end{multline}
The result is shown in \cref{fig:noise budget}, showing that the apparatus is photon shot noise limited over the relevant frequency range.

\section{Cavity splitting}
\label{sec:splitting}

\begin{figure}[t]
    \centering
    \includegraphics[width=\columnwidth]{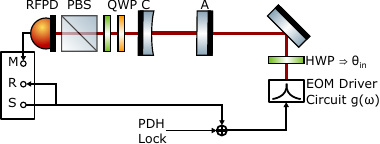}
    \caption{Simplified setup to find cavity splitting and verify experimental calibration. The cavity is represented here only by mirrors A and C for simplicity.
      A network analyzer with a source ``S'' drives the electro-optic phase modulator at the cavity input, generating both p- and s-polarized rf sidebands.
      At the polarimetric readout behind mirror C, one of the p-polarized sidebands beats with the s-polarized carrier field from the cavity, producing a beat note at the rf photodiode.
      The network analyzer measures the transfer function from ``R'' to the beat note signal ``M''.
    }
    \label{fig:calibration}
\end{figure}

To find the $\hat{s}$-$\hat{p}$ cavity splitting $\omega_{\textrm{sp}}$ and verify our calibration, we use the setup shown in \cref{fig:calibration}.
Linearly polarized light, with polarization angle $\theta_{\mathrm{in}}$ controlled by an input half-wave plate, is pumped into the cavity, and the cavity is locked to the $(\omega_0, \hat{s})$ mode. 
The network analyzer is used to scan the sideband frequency $\omega_{\text{sb}}$ for the phase-modulated sidebands driven by the same electro-optic modulator (EOM) used to perform the PDH locking.
When $\omega_{\text{sb}}=\omega_{\textrm{sp}}$, the $(\omega_0 + \omega_{\text{sb}}, \hat{p})$ mode also resonates in the cavity, and we mimic the effect of an ALP-induced phase modulation at that frequency. 
This resonance can be measured at the readout, from which we infer the cavity splitting frequency.

\begin{figure}[t]
    \centering
    \includegraphics[width=\columnwidth]{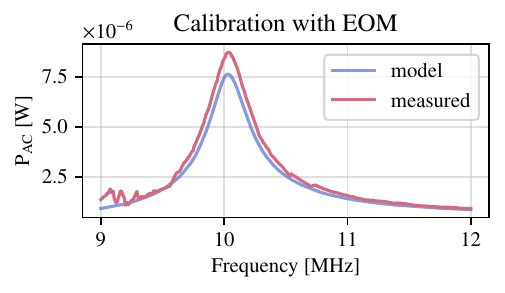}
    \caption{Verification of cavity calibration model (blue) with measurement (pink) for dataset 1.}
    \label{fig:calibration_plot}
\end{figure}

The power in the pump input is $P_0$, and thus the power in the sideband input is given by $P_0 \cos^2{(\theta_{\text{in}}}) \, \eta(\omega)^2/4$, where the modulation depth is defined by
\begin{equation}
    \eta (\omega) = \frac{V_{\text{in}}}{V_\pi} g(\omega),
\end{equation}
where $g(\omega)$ is the transfer function of the EOM driver circuit, $V_{\mathrm{in}}$ is the voltage sent by the network analyzer, and $V_\pi$ is a property of the EOM. 
If $\epsilon$ is the mode-matching and readout efficiency, and $C(\omega)$ is the normalized cavity amplitude transfer function (\cref{eq:Chat_p}), we arrive at a model for the ac power at the rf photodiode:
\begin{equation}
    P_{\mathrm{AC}}^{\mathrm{model}} = 2\sqrt{P_{\mathrm{LO}}P_0} \cos(\theta_{\mathrm{in}}) \lvert C^{\hat{p}}(\omega) \rvert \frac{\eta (\omega)}{2}\epsilon.
\end{equation}
We can then compare our measurement of the transfer function $M/R$ with our model (see \cref{fig:calibration} for the schematic), thus verifying our cavity calibration. 
This transfer function is shown in \cref{fig:calibration_plot} for dataset 1.

\section{Datasets}
\label{sec:datasets}

In Table \ref{tab:datasets}, we provide cavity finesse $\mathcal{F}_s, \mathcal{F}_p$ in $\hat{s}, \hat{p}$ polarizations, and the frequency of peak axion sensitivity given by the $\hat{s}$-$\hat{p}$ cavity splitting $\omega_{\mathrm{sp}}$ for each of our five datasets.

\begin{table}[t]
  \sisetup{table-format=4.2}
  \centering
  \caption{
    Cavity parameters for each dataset.
    The finesse values for the s-polarized pump and p-polarized signal modes are given, along with the central frequency $\omega_{\mathrm{sp}}$ of the signal mode resonance relative to the pump mode resonance.
  \label{tab:datasets}}
  \begin{ruledtabular}
  \begin{tabular}{rSSS}
  Dataset & {$\mathcal{F}_s$} & {$\mathcal{F}_p$} & {$\omega_{\text{sp}}/2\pi$ [MHz]} \\
  \hline
  1 & 7256 & 212 & 10.03 \\
  2 & 6634 & 253 & 10.30 \\
  3 & 7000 & 160 & 12.07 \\
  4 & 5639 & 151 & 13.27 \\
  5 & 6043 & 152 & 13.54 \\
  \end{tabular}
  \end{ruledtabular}
\end{table}

\section{Single photon readout sensitivity}
\label{sec:single_photon}

Instead of a heterodyne readout scheme as performed for this experiment, future axion polarimetry experiments could utilize a single photon readout scheme at the cavity transmission, where one would implement a series of polarizing beam splitters or filter cavities to eliminate any pump photons and make a measurement of whether any signal photons were transmitted from the cavity. Following the ideas laid out in Ref.~\cite{Lamoreaux:2013koa}, the signal would be given by a photon transmission rate of axion generated p-polarized photons
\begin{equation}
    \Dot{N}_p = \frac{P_{\text{out}}^{(\omega_0+\omega_a, \hat{p})}}{\hbar(\omega_0+\omega_a)} \approx \frac{P_{\text{out}}^{(\omega_0+\omega_a, \hat{p})}}{\hbar\omega_0}.
\end{equation}
The single photon detector will have an intrinsic dark count rate $\Dot{N}_{\text{dark}}$. Since there will be a large flow of s-polarized pump field photons transmitted by the cavity, we will also have noise due to a finite extinction ratio of the filter system, given by
\begin{equation}
    \Dot{N}_s = \frac{P_{\text{cav}}^{(\omega_0, \hat{s})}t_C^{\hat{s}} \epsilon}{\hbar\omega_0},
\end{equation}
where $\epsilon$ is the fraction of s-polarized photons that leak out of the filter cavities. To estimate the background noise, the experiment would first be run in a configuration that is insensitive to an ALP signal, for example by using a polarizing beamsplitter to reject p-polarized signal photons.
Assuming Poissonian fluctuations and an integration time $T$, total number of photons would be measured with a mean
\begin{equation}
    \lambda_0 = (\Dot{N}_s + \Dot{N}_d) T
\end{equation}
and with variance $\sigma_0^2 = \lambda_0$.
Then the experiment would be re-run in a configuration that is sensitive to the signal, and total number of photons would be measured with a mean
\begin{equation}
    \lambda_1 = (\Dot{N}_s + \Dot{N}_d+ \Dot{N}_p) T
\end{equation}
and variance $\sigma_1^2 = \lambda_1$.
The presence of the signal could be inferred from a nonzero value for the quantity $\lambda_1-\lambda_0$, which has variance $\sigma_0^2 + \sigma_1^2$.
Expressed as a signal-to-noise ratio, this is
\begin{equation}
    \text{SNR} = \frac{\lambda_1-\lambda_0}{\sqrt{\sigma_1^2+\sigma_0^2}}
    = \frac{\Dot{N}_p T}{\sqrt{\bigl[2\bigl(\Dot{N}_s + \Dot{N}_d\bigr) + \Dot{N}_p\bigr] T}}.
\end{equation}
We do not consider noise due to thermal photons as we are working with terahertz frequency photons, where the thermal occupation can be neglected at room temperature.

\bibliography{adbc.bib}

\end{document}